# Topological Constraint Model of Alkaline Earth Vanadate Glasses

Adam Shearer[*] and John C. Mauro

Department of Materials Science and Engineering, The Pennsylvania State University, University Park, Pennsylvania 16802, USA

* Corresponding author: Adam Shearer, email: aus742@psu.edu

**Abstract**

Topological constraint theory has enabled the successful prediction of glass properties over a wide range of compositions. In this study, a topological constraint model is constructed for alkaline earth vanadate glasses based on experimental data. The change in vanadate structural units from $VO_5$ to $VO_4$ was modeled as a function of alkaline earth content and related to thermal and mechanical properties. The model covers both high and low-temperature properties to probe the temperature dependence of constraint rigidity for each constituent of the glass network. The model is changed to describe anomalies in magnesium sites potentially implying that magnesium can form locally rigid structures. Furthermore, the traditional understanding of vanadate glass structure is compared to recent results concluding that the terminal oxygen must exist as a part of the $VO_4$ units. Results for the model explain that bridging oxygen constraints are the main contributors to network rigidity in both low and high temperature regimes. Vanadate glass networks are highly connected even with the introduction of modifier species, which introduce their own bond constraints. Corroboration between experimental data and the topological constraint model illustrates the role of alkaline earth oxides in the glass network.

1. Introduction

Transition metal oxide glasses have been studied extensively for their unique properties including electrical conductivity or semiconductivity, infrared transparency, and nonlinear optical properties [1–3]. Multivalent transition metals such as vanadium, chromium, manganese, iron, and copper have been investigated not only for the effects on glass properties but also the effects on the glass network structure [4–6]. In a silicate network, these elements can act as glass network modifiers, disrupting the structure by creating nonbridging oxygens. However, vanadium is unique to these elements as it is able to create a glass network by itself.

The structure of vanadate glasses has been examined using many different characterization techniques. Spectroscopy techniques have investigated the change in concentration of V-O bonds versus V=O bonds as a function of compositions, illustrating a change in the structural units that make up the glass network [7,8]. Nuclear magnetic resonance has been employed to investigate different vanadate glasses to probe the coordination environments around different cations to compare to other crystalline structures [9]. Using synchrotron-sourced x-rays and neutrons, the structure of alkali and alkaline earth vanadate glasses by visualizing the radial distribution function experimentally [10–13]. The results of the study by Hoppe et al. show that the average coordination of vanadium atoms in alkali and alkaline earth glasses shifts from 4.5 to 4 with the introduction of modifiers up to 50 mol%. This model suggests that $VO_5$ and $VO_4$ structures exist in equal concentrations in pure $V_2O_5$ glasses. As the modifier is added, the structure shifts entirely to $VO_4$ units at the metavanadate composition.

Efforts to model the structures of vanadate glasses computationally have been made but are difficult due to the lack of classical interatomic potentials for vanadium [14–17]. Topological constraint theory (TCT) is an ideal candidate to evaluate the empirical structural models by employing a simple modeling approach [18,19]. TCT, originally developed by Phillips and Thorpe, examines how the number of atomic degrees of freedom and the number of interatomic constraints affect macroscopic glass properties [20]. Glass properties such as the glass transition temperature ($T_g$), hardness, elastic modulus, among others, can be related to the change in average number of constraints for a given composition [21,22]. By comparing predicted property values to experimental data, a particular structural model can be evaluated.

In this study, TCT is applied to alkaline earth vanadate glasses to predict the glass transition temperature and hardness. Using the structural model from Hoppe et al., the constructed TCT model shows evolution in the glass network rigidity as the concentration of each network-forming species changes [10]. Change in speciation of vanadate structural units drives the change in property values. The TCT model accurately predicts properties at the glass transition temperature and room temperature corroborating the model by Hoppe et al.

2. **Model**

*Structural Model*

To create the topological constraint model, assumptions need to be made about the change in structure as the composition changes. Vanadate glasses have been shown to have two main structural units through various diffraction and spectroscopy techniques. Hoppe et al. have presented high-energy x-ray diffraction and neutron diffraction data illustrating the change in $VO_5$ concentration in the structure as a function of modifier content [10]. The coordination environment of vanadium sites in these glasses is shown in Figure 1, where the average coordination evolves from 4.5 toward 4 with the introduction of modifier. This indicates equal concentrations of $VO_5$ and $VO_4$ in pure $V_2O_5$ glass. As modifier is added up to 50 mol%, $VO_5$ units are converted to $VO_4$. At the $50MO-50V_2O_5$ composition, no more $VO_5$ units will exist. Vanadium atoms may change coordination, but not oxidation state. The five coordinated unit consists of five bridging oxygens (BO), while the four coordinated unit consists of three BOs and one double bonded terminal oxygen (TO). These structural units are illustrated in Figure 1.

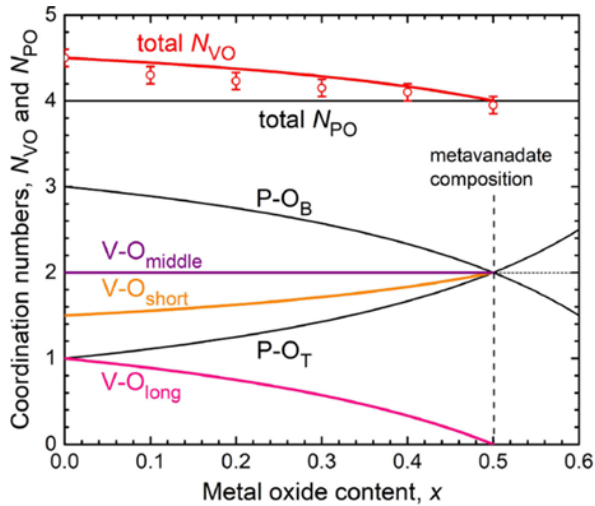 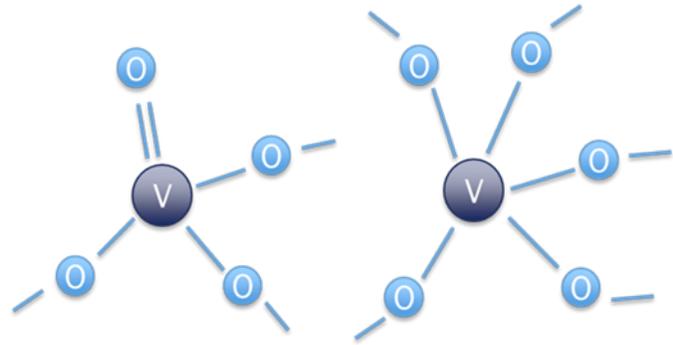

Figure 1: (Left) Coordination environment of vanadium atoms in modified vanadate glasses reproduced from Ref [10]. The top (red) curve illustrating the change in vanadium coordination number from 4.5 to 4 with addition of modifier ions. (Right) Depictions of the vanadate structural units showing the $VO_4$ unit consisting of 3 bridging oxygens and one doubled bonded terminal oxygen. The $VO_5$ unit on the right consists of 5 bridging oxygens connected to other vanadate units.

When developing a structural model for TCT, each individual species contributing to the glass network must be accounted for. Hence, we must consider each individual atomic species rather than oxide units as a whole. Each speciation of vanadium is considered, along with the two oxygen bonding environments, and the modifier atoms. Table 1 lists the species contributing to the glass, along with the designation used for them in this paper, and a description of their bonding environment. The subscript used to describe the different vanadate species in Table 1 indicates the number of bridging oxygens associated with that vanadate unit, not the coordination number of the vanadium. This structural model is employed in this topological constraint model to predict glass transition temperature and glass hardness as a function of composition.

Table 1: Description of each atomic species contributing to the glass network.

| Species | Unit Description |
|---|---|
| $V_5$ | Five-coordinated vanadium atoms in the glass network |
| $V_{3a}$ | Four-coordinated vanadium species with three bridging oxygens and one terminal oxygen. |
| $V_{3b}$ | Four-coordinated vanadium having two bridging oxygens. This species is formed by conversion from the $V_5$ through introduction of modifier |
| BO | Bridging oxygens |
| TO | Terminal oxygens (doubled bonded to a vanadium) |
| M-O | Alkaline earth modifier cations |

*Topological Model*

A glass structural network consists of radial bond stretching (BS) constraints and angular bond bending (BB) constraints that oblige bond length and bond angles to a narrow distribution. The number of radial constraints associated with a particular structural unit is determined by dividing the coordination number of the unit by 2 as each bond is shared by two atoms (Equation 1). For angular constraints, the number is determined by twice the coordination number minus 3, as long as the coordination of the unit is above 2 (Equation 2). Each of these constraints is temperature-dependent, where high thermal energy can overcome the bond constraints, making them "floppy". Weaker units, typically those that are more ionic, can be considered floppy even at lower temperatures and thus do not contribute to the network rigidity. Generally, radial constraints are more likely to be rigid in the glass transition temperature, while angular constraints become rigid at lower temperatures.

$$n_{radial} = \frac{r}{2} \qquad \text{Equation 1}$$

$$n_{angular} = 2r - 3 \qquad \text{Equation 2}$$

The determination of the total number of degrees of freedom for a glass composition is shown in Equation 1, where $3N$ is the initial number of degrees of freedom, $n$ is number of constraints, and 6 corresponds to the total number of macroscopic degrees of freedom.

$$F = 3N - n - 6 \qquad \text{Equation 3}$$

A glass network having exactly zero degrees of freedom is considered to be isostatic where the glass-forming ability is optimized. Fewer than zero degrees of freedom is considered underconstrained, where the network has floppy modes that allow continuous deformation of the glass structure, helping to facilitate crystallization. If the number of degrees of freedom is more than zero, the structure is considered overconstrained, causing rigid structures to percolate through the network again trending towards crystallization. For a covalently bonded network, the degrees of freedom can also be related to the average coordination number of the glass, where an isostatic network typically corresponds to an average coordination number of 2.4.

Using the calculated number of degrees of freedom, TCT can be employed to calculate the glass transition temperature relative to a known reference temperature [23]. Using Equation 3, where $d = 3$ is dimensionality of the glass network and $n[T,x]$ is the average number of rigid constraints at $T_g$, and $n$ is the number of constraints for a given composition [24].

$$\frac{T_g(R)}{T_g(R_r)} = \frac{f[T_g(x_r), x_r]}{f[T_g(x), x]} = \frac{d - n[T_g(x_r), x_r]}{d - n[T_g(x), x]} \qquad \text{Equation 4}$$

Temperature-dependent topological constraint theory also predicts room temperature properties, such as hardness or strength, by considering some combination of radial and angular constraints [22,24,25]. In order to predict the compositional dependence of hardness for a given glass system, there is the requirement that a certain critical number of constraints is required to provide a rigid interconnected structure. When the average number of rigid constraints is less than the critical number, the material will possess no resistance to an incoming force. When the number of rigid constraints per atom is equal to or greater the critical value, the network is able to provide mechanical resistance in the third dimension. A value of $n_{crit}=2$ presents a network that is two-dimensional and rigid in a plane, similar to graphene sheets. A $n_{crit}=3$ value would demonstrate a structure that is rigid in all three dimensions.

The original model developed to predict hardness from constraint models was built considering a purely covalent glass system, such as selenium. Selenium forms 1-dimensional chains with 1 radial constraint and 1 angular constraint per atom. In such covalent glasses, the angles typically provide a high degree of rigidity in the glass networks, sometime more than the radial bond constraints. To resist the force of an indenter, rigidity is needed in three dimensions. For a selenide glass system, the third dimension of rigidity begins when there is more than 2.5 constraints per atom. However, this model rationalized from covalent glasses does not necessarily describe oxide systems where the angular constraints are more weakly defined and radial constraints tend to contribute more to the network rigidity. Two is the minimal number of constraints per atom to form a rigid two-dimensional network. In the absence of angular constraints that fall in the same place as the network of radial constraints, any additional constraints beyond 2 can lead to rigidity in the third dimension. In this work, a critical value of 2 constraints per atom was empirically found to work the best for the alkaline earth vanadate system, as discussed later in this paper. The hardness of a given composition is calculated using Equation 3 where $dH_v/dn$, is a proportionality constant dependent on the load and shape of the indenter [19,26].

$$H_v(x,y) = \left(\frac{dH_v}{dn}\right)[n(x,y) - n_{crit}] \qquad \text{Equation 5}$$

Another approach to determining the hardness of a glass has been proposed by Zheng et al., where the constraint density is considered [26] and applied to silicate and borosilicate glasses. Density and molar mass are considered to convert the total number of rigid constraints per atom to constraint density. Equation 4 illustrates this relationship, where $n'(x)$ is the constraint density, $n(x)$ is the total number of constraints, $\rho(x)$ is the glass density, $N_A$ is Avogadro's number, and $M(x)$ is the molar mass:

$$n'(x) = \frac{n(x)\rho(x)N_A}{M(x)} \qquad \text{Equation 6}$$

3. **Materials and Methods**

*Glass synthesis*

Each glass was batched from reagent grade materials with purity greater than 99.9% including strontium carbonate ($SrCO_3$), barium carbonate ($BaCO_3$), and vanadium pentoxide ($V_2O_5$). The molar formula for each series of glasses is $x$SrO-(100-$x$)$V_2O_5$ or $x$BaO-(100-$x$)$V_2O_5$, where *x*=10,20,30,40 and 50. After batching, powders were shaken in a plastic container for 10 minutes to ensure homogenization. Alumina crucibles were used for melting as some interaction with platinum was noted. All glasses were melted for 1 h at 800 °C. The melt was then poured and rapidly quenched between two aluminum plates.

*X-ray diffraction*

The non-crystallinity of the glasses was confirmed through X-ray diffraction (XRD) using a PANalytical Empyrean 1 (Malvern PANalytical Inc., Westborough, MA, USA). XRD was completed over the 2θ range of 5°-70° 2θ with a Cu-Kα X-ray source, and a step size of 0.026°.

*Electro-probe micro-analysis*

Major element concentrations were determined using a Cameca SXFive Electron Probe Micro-Analyzer (EPMA) located in the Materials Characterization Laboratory at Penn State. The instrument is equipped with five wavelength dispersive spectrometers and a LaB6 electron source. An accelerating voltage of 15 keV and a beam current of 30 nA were used. The produced X-ray intensities were subject to a PAP (phi-rho-z) matrix correction algorithm as described by Pouchou and Pichoir and converted to concentrations by comparison to Corning glass and metal oxide standards [27].

*Differential scanning calorimetry*

Glass transition temperature of each glass was determined using differential scanning calorimetry (DSC). The samples were ground into a fine powder and roughly 20 mg were loaded into an alumina pan for measurement. The powder was annealed within the DSC to erase the thermal history. The glass transition was measured with a ramp rate of 10 °C per minute up to 400 °C. The onset method was used to determine the $T_g$ [28].

*He-Pycnometry*

The glass density was measured using a AccuPyc™ II 1340 (Micromeritics, Norcross, GA) through gas displacement. A sample size of roughly 1 g of sample was measured 10 times to minimize instrumental error.

*Microindentation*

Glass samples were annealed just below the corresponding measured glass transition temperatures (0.92$T_g$) for 6 hours and allowed cool to room temperature. The annealed glass samples were mounted in epoxy and polished to a surface roughness of <0.25µm. Polishing was completed using a 3-sample MetPrep 4 automatic polisher (Allied High Tech Products Inc., Compton, CA, USA). Silicon carbide grinding pads and white label cloth pads with poly-diamond suspensions were used to polish to the desired surface roughness.

Microindentation was performed using a Q60A+ automated microindenter (Qness GmbH). The Vickers hardness of all the glass samples was determined with a minimum of 10 indentations conducted under ambient conditions. Each indentation was held for 25 s, in accordance with ASTM C1327-15(2019). The Vickers microhardness ($H_V$) in GPa was calculated using Equation 7, where the load $P$ is in N and $d$ is the average diagonal length in mm [29].

$$H_v = 0.0018544 \cdot \left(\frac{P}{d^2}\right) \qquad \text{Equation 7}$$

**Results**

*Glass Confirmation*

To ensure that the chemical composition and non-crystalline nature of the glasses were preserved, EPMA and XRD were employed respectively. EPMA provides the evaluated chemical compositions and is compared to the batched compositions in wt% shown in Table 2. Figure 2 provides the diffractograms for both series of glasses confirming that there is no crystallinity to the samples.

Table 2: Batched and measured compositions of strontium and barium vanadate glasses using EPMA along with measured density values.

| Sample | Batched Composition [wt%] | | EPMA Results [wt%] | | | Density [g/cm³] |
|---|---|---|---|---|---|---|
| | MO | $V_2O_5$ | MO | $V_2O_5$ | $Al_2O_3$ | |
| **Sr10** | 5.95 | 94.05 | 6.62 | 92.37 | 0.59 | 3.27 |
| *Sr20* | 12.47 | 87.53 | 13.67 | 85.59 | 0.85 | 3.36 |
| *Sr30* | 19.62 | 80.38 | 17.65 | 80.48 | 0.31 | 3.42 |
| *Sr40* | 27.53 | 72.47 | 24.13 | 73.43 | 0.12 | 3.52 |
| *Sr50* | 36.29 | 63.71 | 32.26 | 63.25 | 0.11 | 3.79 |
| *Ba10* | 8.57 | 91.43 | 8.94 | 89.98 | 1.08 | 3.06 |
| *Ba20* | 17.41 | 82.59 | 17.83 | 81.70 | 0.47 | 3.23 |
| *Ba30* | 26.54 | 73.46 | 26.33 | 74.96 | 0.46 | 3.53 |
| *Ba40* | 35.98 | 64.02 | 35.34 | 65.33 | 0.41 | 3.72 |
| *Ba50* | 45.74 | 54.26 | 44.96 | 55.93 | 0.40 | 3.90 |

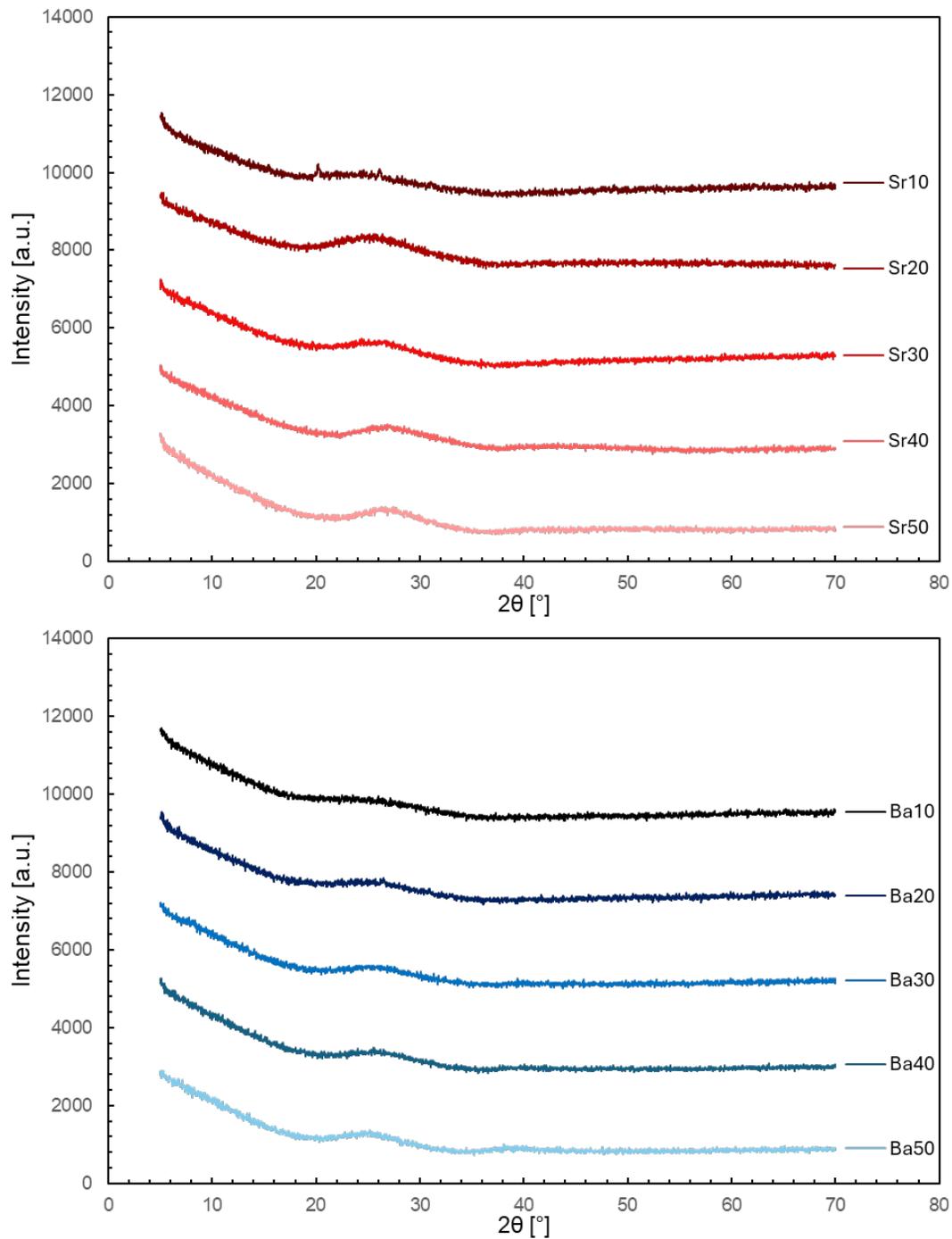

Figure 2: Diffraction patterns for strontium and barium vanadate glasses.

*Network Forming Species*

Each species must be considered when calculating the connectivity of atoms in a glass network,. For the system considered in this paper, the glass network can contain two different vanadate structural units, where $VO_5$ is the 5 coordinated structural unit and $VO_4$ is 4 coordinated with 1 doubly-bonded terminal oxygen (TO). The oxygen in the network can exist as bridging oxygen (BO) that connect different structural units or terminal oxygens that are double bonded with a vanadium atom. The BOs that connect alkaline earth atoms to vanadium atoms are considered separately in this model to better determine their effects on the glass network.

As the concentration of alkaline earth modifiers increases, the primary change can be seen in bridging oxygen concentration. The addition of modifier lowers the overall atomic concentration of oxygen in the glass as well as shifts the oxygen speciation. Converting $VO_5$ units to $VO_4$ removes two bridging oxygens to create one terminal, double bonded oxygen. Concentration changes between the speciation of vanadium, bonding environment of oxygen, and amount of alkaline earth are used to determine the average number of constraints and thus related to macroscopic property changes. Figure 3 highlights the changes in concentration for each glass network forming species as a function of alkaline earth concentration.

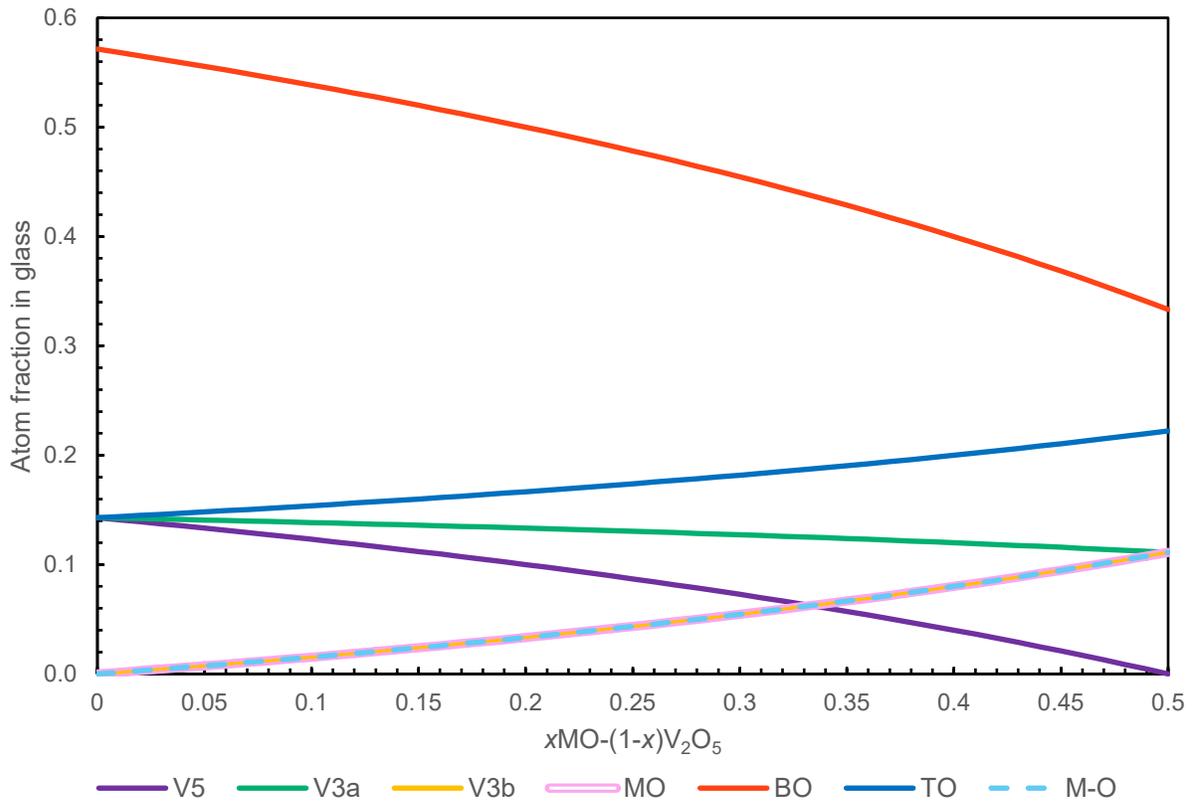

Figure 3: Network forming species changes as the function of alkaline earth concentration.

*Property Predictions from Model*

*Glass Transition*

The temperature dependent TCT model was developed to determine the effect of thermal energy on constraints to enable prediction of properties at nonzero temperatures. When there is enough thermal energy to break a constraint, that constraint no longer contributes rigidity to the glass network. Each constraint mode will break at different temperatures and thus contribute differently to the rigidity of the glass. Predicting properties such as the glass transition temperature must consider the constraints that are rigid at those temperatures of interest.

Figure 4 shows glass transition values from literature for various alkaline earth vanadate glasses along with experimental results from this work. The plotted curve is the predicted glass transition temperature for the alkaline earth vanadate system using topological constraint theory. Experimental values from Basu et al are shown for alkaline earth vanadate glasses containing MgO, CaO, SrO, and BaO [30]. Each of these series follows the trend of the model. However, the magnesium vanadate series does deviate from the prediction towards higher values. At lower modifier contents (below 10 mol%) there is also deviation from the model, which could be attributed to the formation of octahedral vanadate sites as described in Hoppe et al. [10]. An octahedral site vanadate structure would be overconstrained, which would in turn increase the glass transition temperature of the system. However, it is also possible that the constraint onset temperature is lower than the glass transition temperature rendering it floppy at high thermal energies. Another consideration is that at lower modifier content it may be thermodynamically favorable to form $V^{4+}$ sites which can contribute differently the average number of constraints in the glass.

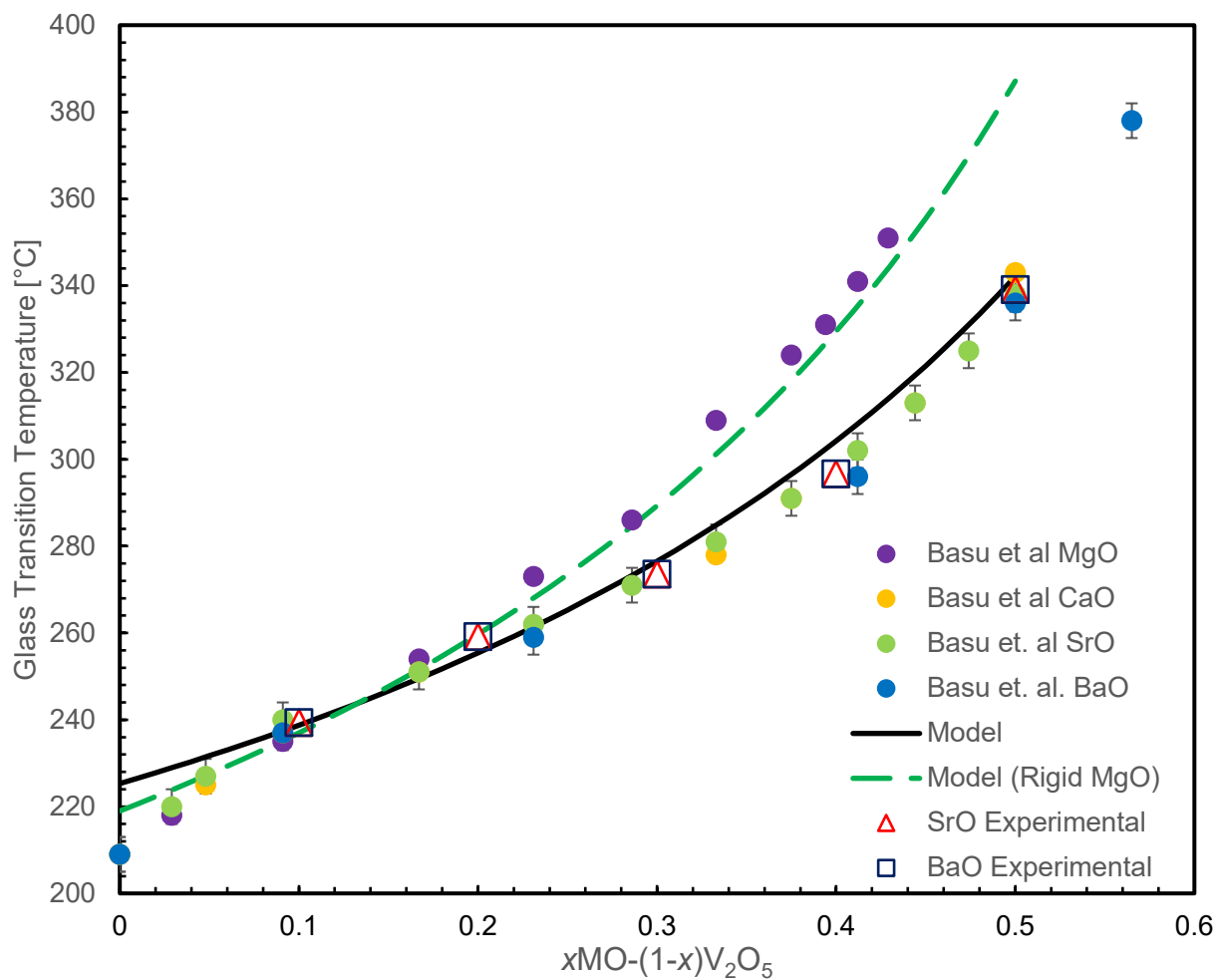

Figure 4: Predicted and experimental glass transition temperature of alkaline earth vanadate glasses. The solid line shows predicted property values from the model while markers are experimental values from literature (filled circles) or this study (open symbols). The green dashed curve represents the adjusted model representing the modifier as locally rigid structures.

The magnesium series of glass transition temperatures are notably higher than those of similar composition and are higher than the temperature predicted through the topological constraint model. Magnesium has a much cation higher field strength than other alkaline earth species and is known to show similar effects in properties of other glass systems providing one description of why the glass transition temperature is higher than other compositions [31–35]. Some studies have previously discussed magnesium as a potential glass network former through experimental evidence showing $MgO_4$ and $MgO_5$ structural units [36–40]. These studies typically contain very high amounts of magnesium. This is a possible explanation to why magnesium vanadates exceed the predicted glass transition temperature. When the model is adjusted to treat magnesium as a tetrahedral unit, the predicted glass transition temperature fits the experimental values for magnesium vanadates from literature.

*Glass Hardness*

As previously mentioned, at room temperature some combination of radial and angular constraints are active and contributing towards the glass rigidity. Some models consider only angular constraints while others consider a combination of both radial and angular constraints [25,26]. This model considers all constraint modes to be rigid with the exception of the $VO_5$ angular constraints. This follows the work of another $V_2O_5$ containing TCT model, where a combination for some species of vanadium both radial and angular constraints will be rigid [41]. The traditional approach for predicting trends in glass hardness utilizes a $n_{crit}$=2.5 value which provides rigidity in two dimensions in a glass structure with partial rigidity in the third dimension. However, this value for $n_{crit}$ is too high for the alkaline earth vanadate glass system. As the total number of rigid constraints in a glass system approaches 2.5, the slope for predicted hardness becomes too steep. Thus, for this study $n_{crit}$=2 is used as the minimum number of rigid constraints for a system to display resistance to an incoming force. Figure 5 illustrates the prediction differences in glass hardness using different values for $n_{crit}$.

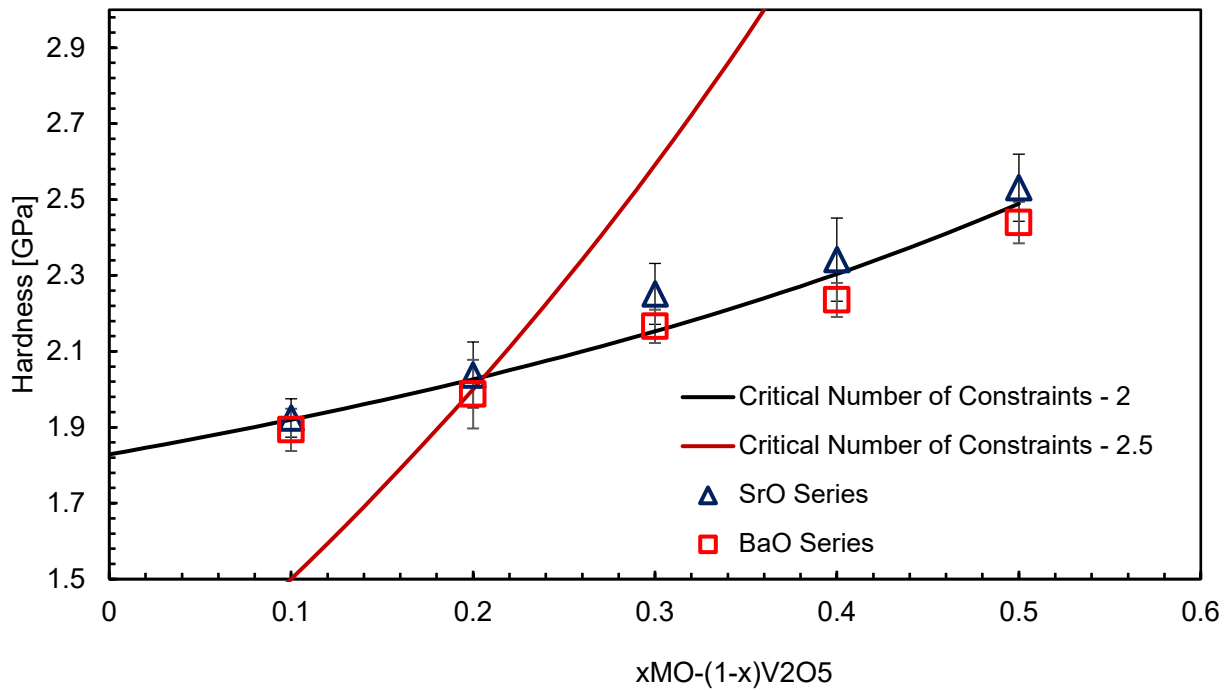

Figure 5: Demonstration in difference of glass hardness prediction between different values of critical number of constraints.

In order to accurately predict the hardness values for this system, a fitting parameter is derived from the slope of experimental hardness values against active constraint modes. This fitting parameter is shown in Figure 6A and is used as the coefficient in Equation 3 to predict hardness for the compositional system. Figure 6B illustrates the predicted hardness for the alkaline earth vanadate system using TCT as well as constraint density theory with experimental values. The linear prediction of glass hardness using constraint density theory does not fit as accurately as the traditional approach using a $n_{crit}$ value of 2.

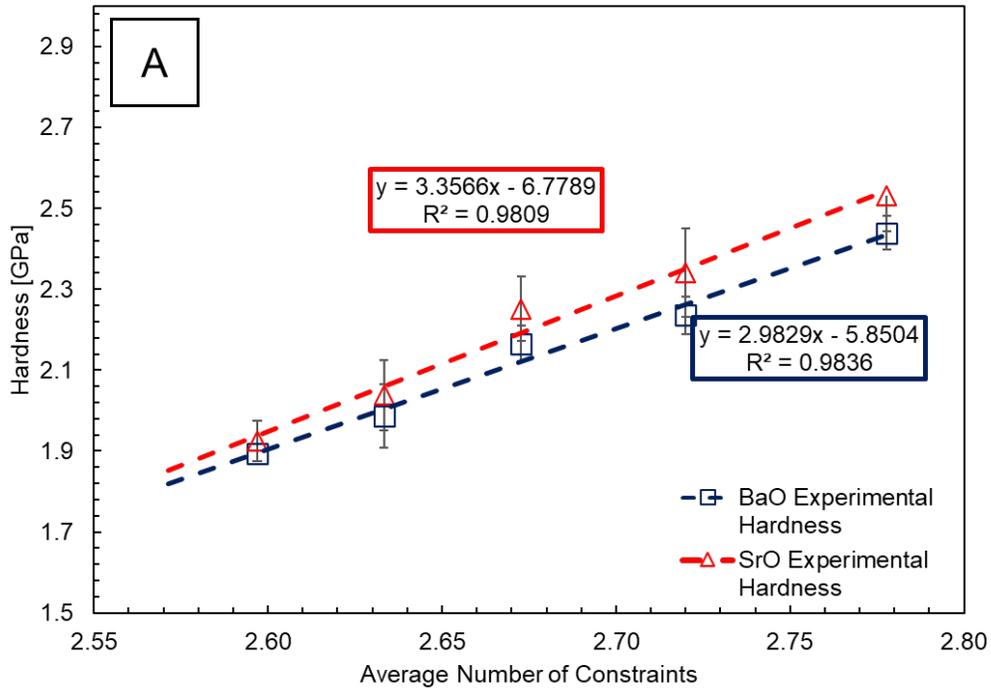

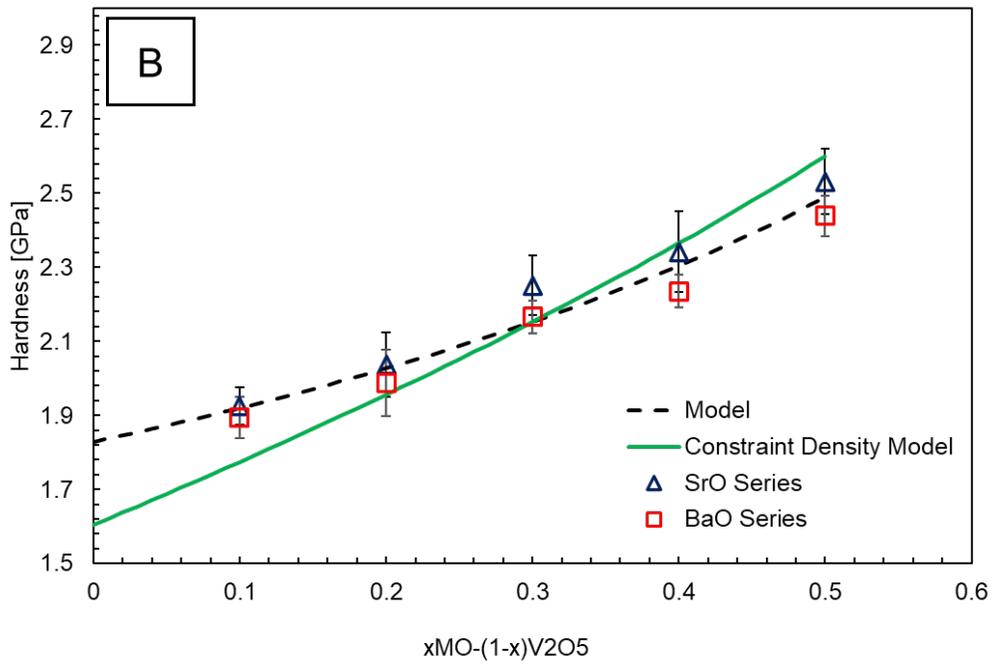

Figure 6: (A) Linear fit of hardness as a function of rigid constraints. (B) Predicted hardness values for alkaline earth vanadate glasses. The solid black line shows the predicted hardness using TCT while the dashed green line shows predicted values using the constraint density method.

## 4. Discussion

The model proposed by Hoppe et al. predicts the speciation of vanadium as a function of the increasing modifier content from x = 0 to 50 mol% [10]. In the absence of modifier molecules, the configuration of vanadium species is shown to have equal concentrations of $VO_4$ and $VO_5$ units. The findings from Hoppe et al.'s study were crucial in developing the topological model presented in this work. The close match between the experimental data and the predicted model validates the accuracy of this vanadate speciation and coordination.

There are contrary reports in literature concerning the structural units of vanadate glasses. Traditionally, vanadate glasses are believed to consist of the two structural units adopted in this study, where a $VO_5$ unit has 5 bridging oxygens and a $VO_4$ unit consists of 3 bridging oxygen and 1 terminal oxygen. This model has been supported by studies over previous decades using various techniques including x-ray diffraction, Raman spectroscopy, and nuclear magnetic resonance [13,30,42–47]. However, the recent study by Hoppe et al indicates that the terminal oxygen may exist on the $VO_5$ unit rather than the $VO_4$ [10]. The authors describe the structural units as similar to the cubic $V_2O_5$ crystal structure. To illustrate these changes, Figure 7 provides visualization of the model used by this study and the new structures proposed by Hoppe et al [10].

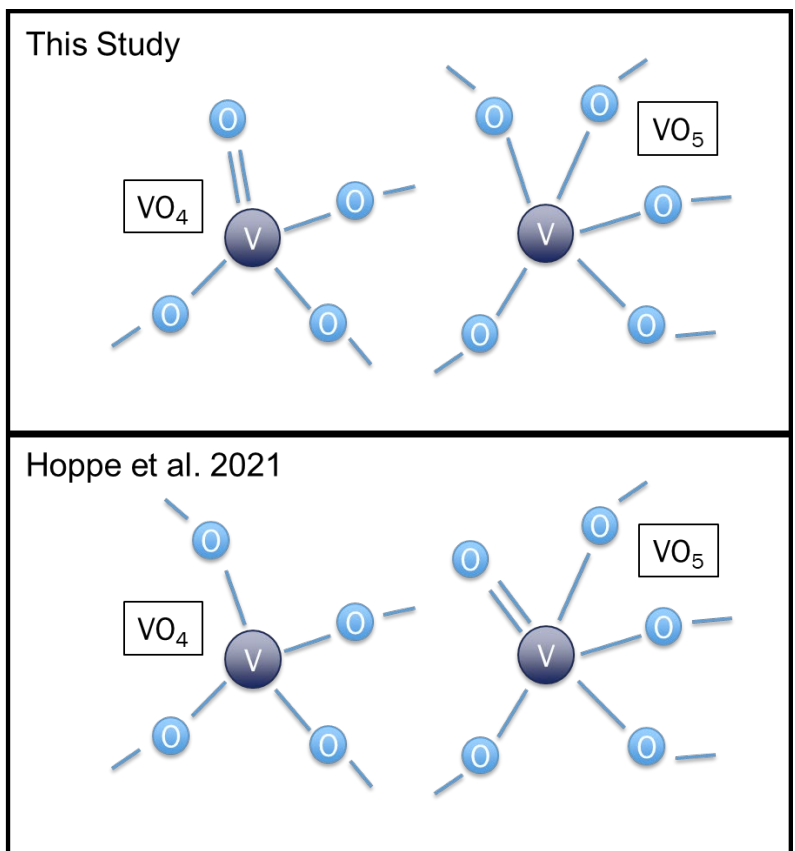

Figure 7: (Top) Structural units for vanadate glasses used in this study based on previous work. (Bottom) Structural units for vanadate glasses as proposed by Hoppe et al. [10].

To understand the effects of this alternative in Figure 7(b) structural model, our TCT model was adjusted to consider the speciation and constraint counting of this alternative model. The alternative model overpredicts the glass transition temperature significantly as shown in Figure 8. The conclusion that the structural model proposed by Hoppe et al. fits the magnesium series of glass can be made but is unlikely the case [10]. Magnesium is not likely to change the structure of the glass this significantly as it does not behave in such a manner in any other glass system. Magnesium is known to exhibit anomalies in properties as compared to other alkaline earth elements due to field strength effects or through potentially forming locally rigid structures around the magnesium site. The argument that the terminal oxygen exists on the $VO_4$ unit is further backed through Raman spectroscopy results. Peak assignment place the V=O bond around a Raman shift of $900 - 1000$ cm$^{-1}$, depending on the system, which is shown to increase in intensity significantly with increasing modifier concentration [7,8,45,46]. Hoppe et al (2021) shows that VO$_4$ concentration also increases with modifier concentration [10]. The conclusion can be made that the VO$_4$ unit is linked to the increase in V=O concentration, therefore the terminal oxygen is attached to the VO$_4$ structure.

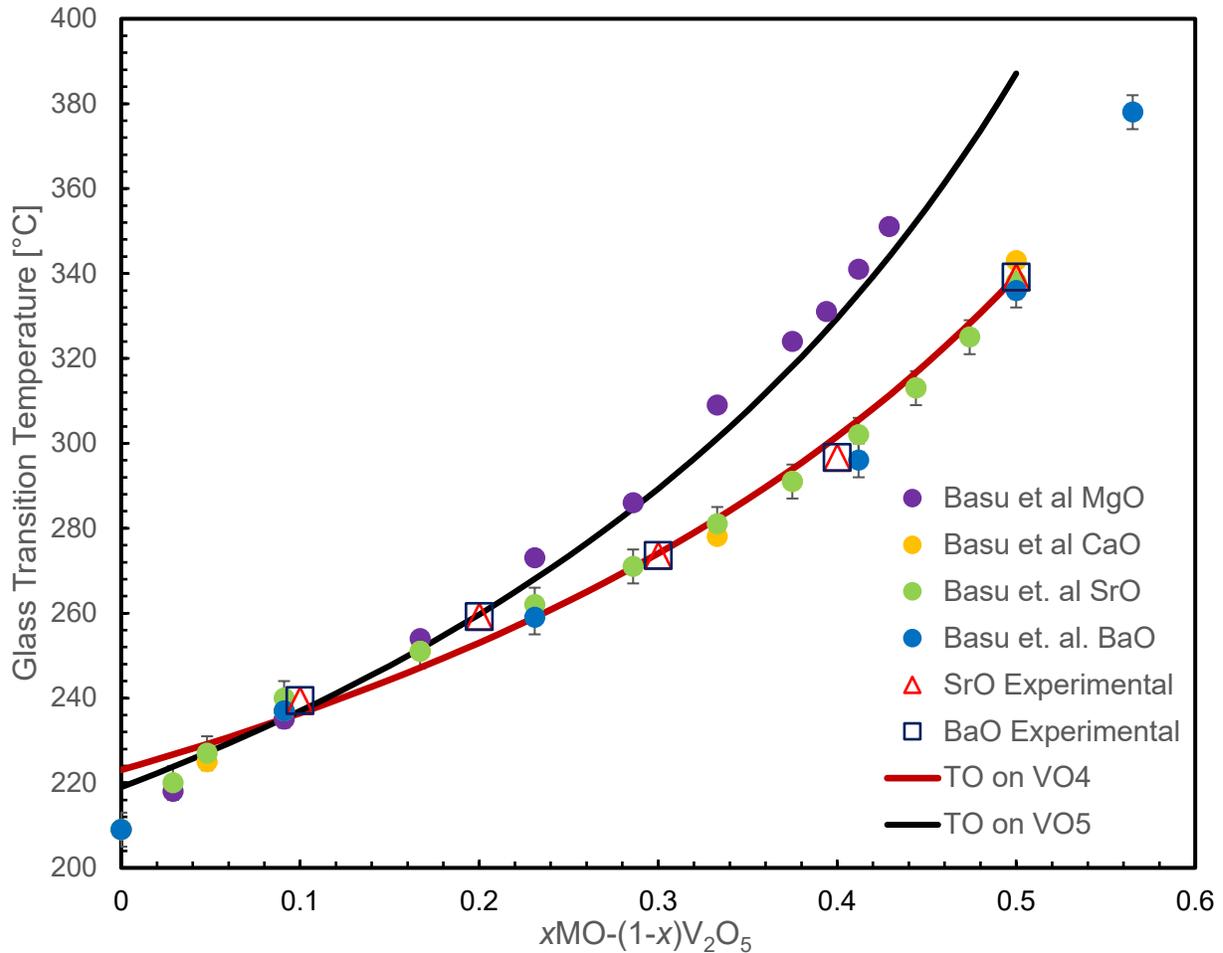

Figure 8: Evaluation of the glass transition temperature predictions by changing the location of the terminal oxygen between the VO$_4$ and the VO$_5$ structures.

Although many characterization techniques have evaluated the structure of different vanadate glass systems, there has yet to be a conclusive model describing their structural evolution. Many systems are complex and difficult to deconstruct into individual structural units. TCT allows for a simplification of each component of the composition to provide a new perspective into the underlying changes. As the composition of the glass changes, concentration of each structural unit is accounted for and allows for the accurate prediction of glass properties. The interpretation of MO-$V_2O_5$ structures allows for a stronger understanding of the evolution of the glass system and is corroborated by the fitting of experimental property results. Through room temperature and elevated temperature property measurements, the argument for the proposed model is strengthened. The temperature dependence of each constraint is shown in Table 3 where the rigid and floppy modes are compared at room temperature and the glass transition temperature. Rigid and floppy constraint modes are similar to those discussed in Shearer and Mauro for a modified $V_2O_5$-$TeO_2$ glass system [41].

Table 3: Temperature dependence of each structural unit constraint modes at room temperature and the glass transition temperature. The number of constraints for the given rigid constraint mode in shown in parentheses.

|  | Room Temperature | | Glass Transition Temperature ($T_g$) | |
|---|---|---|---|---|
|  | **Radial** | **Angular** | **Radial** | **Angular** |
| $V_5$ | Rigid (2.5) | Floppy | Rigid (2.5) | Floppy |
| $V_{3a}$ | Rigid (2) | Rigid (5) | Rigid (2) | Floppy |
| $V3b$ | Rigid (2) | Rigid (5) | Floppy | Rigid (5) |
| MO | Rigid (1) | Rigid (1) | Rigid (1) | Rigid (1) |
| BO | Rigid (1) | Rigid (1) | Rigid (1) | Floppy |
| TO | Rigid (0.5) | N/A | Rigid (0.5) | N/A |
| M-O | Rigid (1) | Rigid (1) | Rigid (1) | Floppy |

Using temperature dependent topological constraint theory, we can determine the contribution of each structural unit towards the properties of interest. To calculate this, the number of constraints for each individual unit is divided by the overall number of constraints, and then multiplied by the either the glass transition or hardness property value. In the case of alkaline earth vanadate glasses, the bridging oxygens contribute the most towards glass transition temperature as they have two radial constraints, more than the other oxygen species and the modifier cation constraints. The vanadate species listed are $V_5$ for five coordinated units, $V_{3a}$ (naturally occurring four coordinated units, and $V_{3b}$ for converted four coordinated units). Figure 9 highlights the contribution of each constraint on glass transition temperature and hardness.

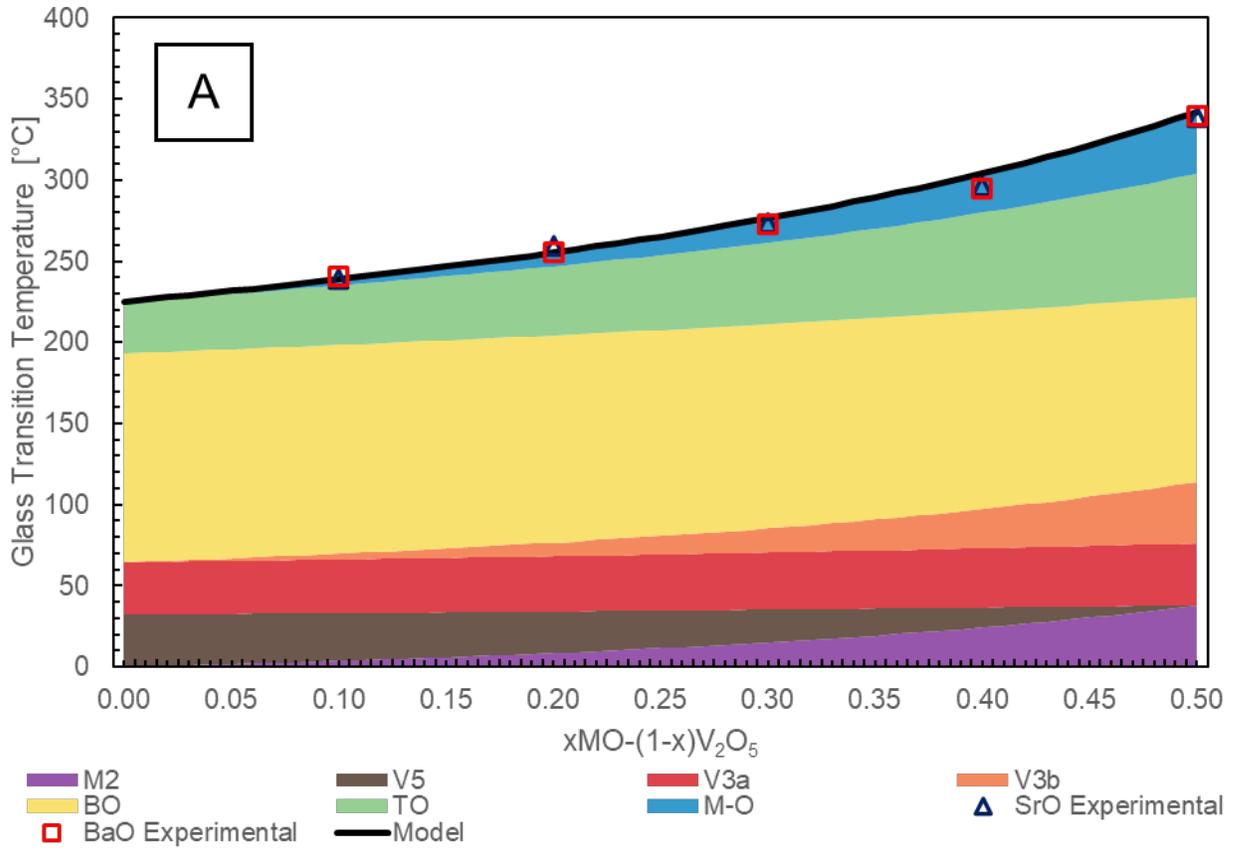
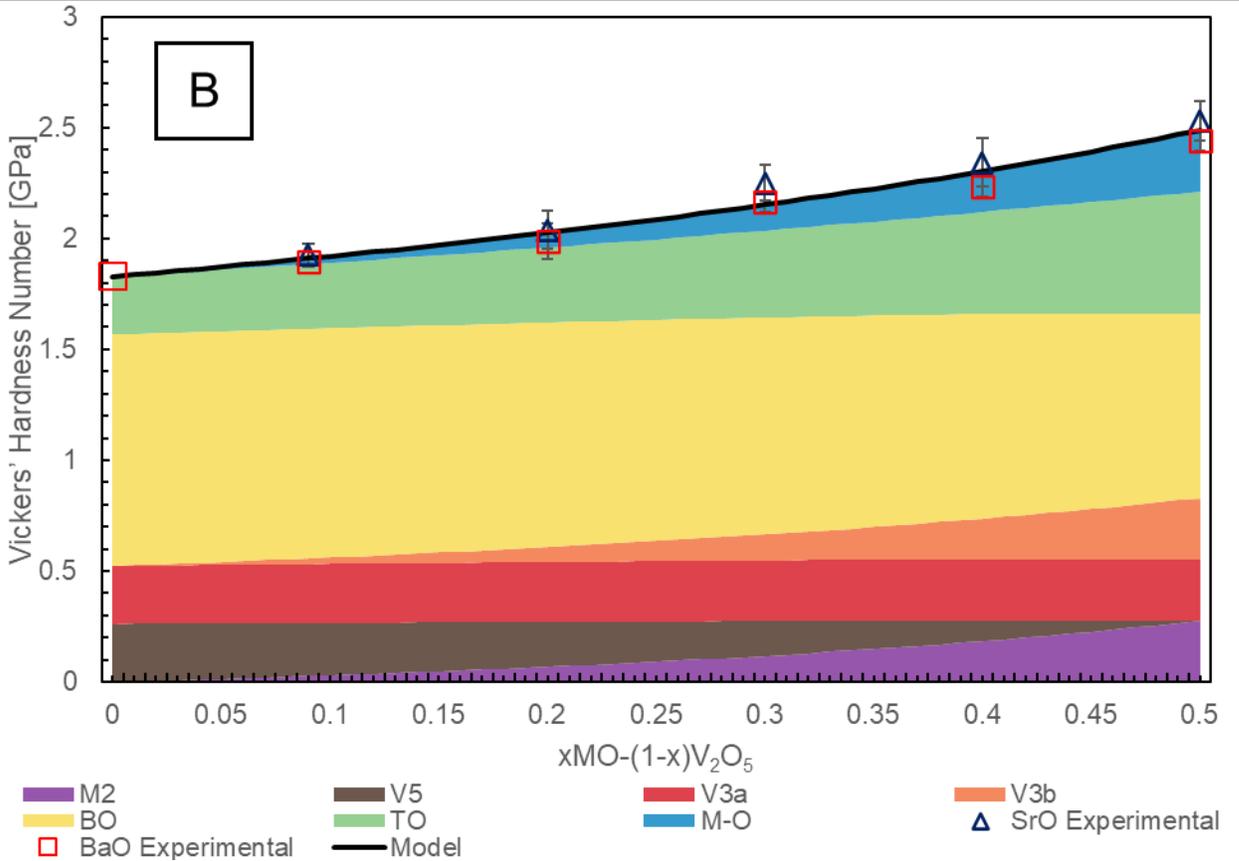

Figure 9: Compositional dependence of glass transition and hardness properties predicted by TCT highlighting the contribution of each structural unit. Individual constraint contributions were calculated by multiplying the predicted property by the number of rigid constraints for the structural unit. Then, the number was divided by the total number of rigid constraints. (A) Predicted glass transition temperature of the compositional workspace highlighting experimentally characterized values. (B) Predicted hardness values at room temperature as a function of composition. Error bars for the glass transition temperature are smaller than the data markers while hardness error bars are one standard deviation from the mean.

The topological constraint model proposed in this manuscript is shown to accurately predict roperties at room temperature and the glass transition temperature. The structural units of alkali vanadate glasses proposed by Hoppe et al. was used develop the concentration of each individual glass forming species. Furthermore, existing publications reporting the temperature dependence of constraints aided the assignment of rigid versus active modes. The accurate prediction of these properties shows that these glasses can be simplified into five coordinated $VO_5$ and $VO_4$ units. Understanding the coordination changes as a function of modifier will assist in the future development of vanadate glasses enabling advances in applications of glasses in energy storage, infrared transparency, nonlinear optical processes, and electronic conduction. Furthermore, adjusting the topological constraint model to fit magnesium vanadates with the assumption that magnesium forms tetrahedral units fits the experimental values from literature. Although no experiments probe the coordination of magnesium in vanadate glasses, this model provides further insight into the possible glass network forming ability of magnesium depending on the chemistry of the glass.

The application of TCT to model vanadate glasses represents a significant advancement in our ability to understand and manipulate the structural and functional properties of these materials. By providing a framework to predict and control the atomic network constraints, this modeling method not only enhances our comprehension of the fundamental behaviors of vanadate glasses but also paves the way for the development of materials with tailored properties. Furthermore, understanding these structural changes allow for the optimization of glass forming ability for realistic glass synthesis in a manufacturing setting. The broader impacts of this approach extend across multiple domains, from improving the performance of optical devices and sensors to advancing the design of durable and high-strength glass materials for various industrial applications [43,48–51].

## 5. Conclusions

This study investigates the fundamental structural units of vanadate glasses modified by alkaline earth oxides through structure-property relationships using topological constraint theory. By using preexisting publications providing insight on structural changes in different vanadate systems, the TCT model was developed with assignment for the temperature dependence of the constraints. This model predicts the glass transition temperature and hardness of the $x$MO-(1-$x$)V$_2$O$_5$ with accuracy within the error bars of the experimental data. Additionally, the model fits experimental results of glass transition temperature of various alkaline earth vanadates from literature, with the exception of magnesium. Treating magnesium as a tetrahedral site improved the prediction of the model for this series, fitting the experimental data. Furthermore, two different proposed structural models are compared using topological constraint theory. Comparison of the topological models to experimental property data indicate that the terminal oxygen exists on the VO$_4$ rather than VO$_5$ structural unit, corroborating the conventional understanding of vanadate glass structure. Understanding the structure of the alkaline earth vanadate family allows for more opportunities to study this family in applied settings including infrared optics, glassy semiconductors, or in cathode materials for fuel cells.

**Acknowledgments**

This research was supported by the Office of Naval Research under grant number N00014-22-1-2590.